\documentclass[12pt]{article}
\usepackage{epsfig}

\begin{document}
\title{A numerical study of the Schr\"odinger-Newton equation \newline
 1: Perturbing the spherically-symmetric stationary states}
\author{R Harrison, I Moroz and K P Tod \\ Mathematical Institute \\ St Giles \\ Oxford OX1
3LB \\\ }


\maketitle

\begin{abstract}

\noindent 

In this article we consider the linear stability of the spherically-symmetric
stationary solutions of the Schr\"odinger-Newton equations. These have 
been found numerically by Moroz et al
\cite{cq98} and Bernstein et al \cite{mp98}. The ground state,
characterised as the state of lowest energy, turns out to be linearly
stable, with only imaginary eigenvalues. The $(n+1)$-th state is 
linearly unstable 
having $n$ quadruples of complex 
eigenvalues (as well as imaginary eigenvalues), where a quadruple consists of
$\{\lambda,\bar{\lambda},-\lambda,-\bar{\lambda}\}$ for complex $\lambda$.

\end{abstract}

\section{Introduction}

The Schr\"odinger-Newton equations (hereafter the SN
equations) for a quantum-mechanical particle of mass $m$ moving in its
own gravitational potential, are the pair of coupled non-linear 
partial differential equations:
%
\begin{eqnarray}
i\hbar{}\frac{\partial\Psi}{\partial{}t} &=&
-\frac{\hbar^2}{2m}\nabla^2\Psi + m\Phi\Psi  \label{SNE1}\\
\nabla^2\Phi &=& 4\pi{}Gm|\Psi|^2 \nonumber
\end{eqnarray}
%
where $\hbar$ is Planck's constant, $\Psi$ is the wave function, 
$\Phi$ is the gravitational potential, $G$ is 
Newton's gravitational constant and $t$ is time. 

The SN equations were introduced by Penrose 
\cite{gr96},\cite{pt98} in his discussion of quantum state reduction
by gravity. Penrose suggested that a superposition of two quantum
states corresponding to two separated `lumps' of matter
 should spontaneously reduce 
to one or other of the states within a time related to the self-energy 
of the gravitational field generated by the difference of the densities. 
This idea requires that there be a preferred basis or special 
set of quantum states which
collapse no further and these, Penrose suggested, should in the
non-relativistic limit be the
stationary states of the SN equations.

The spherically-symmetric stationary states of the SN equations have been found
numerically by a number of authors starting with Ruffini and Bonazzola 
\cite{pr69} and including  Moroz, Penrose and Tod \cite{cq98} and 
Bernstein, Giladi and Jones \cite{mp98}. Our aim in this article is to 
test the linear stability of
these solutions by 
linearising the time-dependent SN equations about a
spherically-symmetric stationary state. In a later article, we 
shall present a numerical
study of the full non-linear evolution in certain symmetric cases.

The SN equations as introduced by Penrose are closely related to the
Schr\"odinger-Poisson equations which have been studied for much
longer (see e.g. \cite{rmp95}). They are also the non-relativistic
limit of the Einstein equations with a complex Klein-Gordon field as
source (see e.g. \cite{pr69}) which is why their solutions are
sometimes known as boson stars (see e.g. \cite{sb95}).\\

The plan of this article will be as follows. In the remainder of this
section we review the nondimensionalised SN equations and collect some
analytical results on them. In Section 2, we review the
spherically-symmetric stationary states and in Section 3 we derive the
linearised time-dependent SN equations. In Section 4 we 
describe the numerical methods
which will be used to solve the linearised equations and in Section 5
we present the results. 

We shall find that the ground-state 
solution is linearly stable,
having purely imaginary eigenvalues, and the $n$-th excited state,
equivalently the $(n+1)$-th state, has
$n$ unstable modes corresponding to $n$ quadruples of complex
eigenvalues, together with purely imaginary eigenvalues. These results
are in line with the expectation one might have
from similar nonlinear problems like those described in \cite{volk}
and \cite{biz}. \\

Following Moroz et al \cite{cq98} we begin by introducing dimensionless variables for the system 
(\ref{SNE1}). We write the 
rescaled $\Psi, \Phi$ as $\psi, \phi$, and we
rescale space and time but write them using the same 
variables as before to obtain the
nondimensional SN equations:
\begin{eqnarray}
i\frac{\partial\psi}{\partial{}t} &=& -\nabla^2\psi +\phi\psi,\label{ndimS}\\
\nabla^2\phi &=& |\psi|^2. \nonumber
\end{eqnarray}
The most important rescaling, which we note for later use, is that for
 the time variable, for which 
$t_{new}= \gamma t_{old}$ with
\begin{equation}
\gamma = \frac{32\pi^2G^2m^5}{\hbar^3}. \label{gamma}
\end{equation}
Stationary solutions as usual take the form $\psi(x,y,z)e^{-iEt}$ 
and satisfy the nondimensional time-independent SN
equations:
\begin{eqnarray}
E\psi &=& -\nabla^2\psi + \phi\psi,  \label{ntdSNa}\\
\nabla^2\phi &=& |\psi|^2. \label{ntdSNb}
\end{eqnarray}
At this point it is worth collecting some analytical results on 
the SN equations. 
\begin{itemize}
\item Existence and uniqueness for 
solutions of (\ref{ndimS}) is established by the following
simplified version of a theorem of Illner et al \cite{Illner}.
\begin{quote}
	Given $\chi(x) \in H^2(\mathbf{R}^3)$ with $L_2$-norm equal to
	1, the system (\ref{ndimS}) has a unique, strong solution
	$\psi(x,t)$, global in time, with $\psi(x,0) = \chi(x)$ and
	$\int|\psi|^2 = 1$. 
\end{quote}
Illner et al \cite{Illner} give precise regularity 
properties for the resulting solution.
\item The equation (\ref{ntdSNa}) can be 
obtained as a variational problem from
the Lagrangian:
\begin{eqnarray}
{\mathcal{E}}=
\int(\nabla\psi\cdot\nabla\bar{\psi}+\frac{1}{2}\phi|\psi|^2) 
\label{act}
\end{eqnarray}
where it is understood that $\phi$ is the solution of (\ref{ntdSNb}), 
and the overbar denotes complex conjugation. 
(All integrals will be over $\bf{R^3}$ unless indicated to the
contrary.)
 Here $E$ is a Lagrange multiplier for the normalisation constraint
\begin{equation}
 I = \int{}|\psi|^2 = 1.
\label{I}
\end{equation}
\item The quantity $\mathcal{E}$ of (\ref{act}) is 
bounded below (\cite{Tod2}). It is
reasonable to suppose, though a proof has not appeared in the literature,
that the infimum of $\mathcal{E}$ is attained and 
that the ground state is the lowest
energy spherically-symmetric solution found numerically by the authors
mentioned above and proved to exist in \cite{mt98}.

\item The system (\ref{ndimS}) preserves the normalisation (\ref{I})
  and the quantity ${\mathcal{E}}$ of (\ref{act}) even with
  time-dependent $\psi$ and $\phi$. We shall therefore 
call ${\mathcal{E}}$ the
  conserved energy. Note that it is different from the energy
  eigenvalue $E$ for a stationary state appearing in (\ref{ntdSNa}). \\\

\item If we define the kinetic energy $T$ and potential energy $V$ in the
  obvious way by
\begin{equation}
T = \int|\nabla\psi|^2, \,\,V = \int\phi|\psi|^2, 
\label{TV}
\end{equation}
then ${\mathcal{E}} = T+\frac{1}{2}V$. It was shown in \cite{Tod2} 
that, in a stationary state,
\begin{equation}
T = -\frac{1}{3}E, \,\,
V = \frac{4}{3}E, \,\,
{\mathcal{E}} = \frac{1}{3}E. 
\label{TVE}
\end{equation}
In particular, as one expects, the energy eigen-values are all negative.\\\

\item There is an interesting observation on the dispersion of the
wave-function due to Arriola and Solar \cite{as99}. Define the 
second moment $Q$ by
\begin{eqnarray}
Q = \int|\bf{x}|^2|\psi|^2
\end{eqnarray}
assuming that the initial data is such that this integral exists, 
then it follows from (\ref{ndimS}) that
\begin{eqnarray}
\ddot{Q} = \int(8|\nabla\psi|^2 + 2\phi|\psi|^2) =8{\mathcal{E}}-2V.
\end{eqnarray}
Now recall that, as a consequence of the maximum principle (see
e.g. \cite{Max}) $\phi$ is everywhere negative and therefore so is
$V$. Thus if the conserved energy $\mathcal{E}$ is positive, then the
dispersion grows at least quadratically in time. In this sense, a
positive energy solution necessarily scatters.
\end{itemize}

\section{The spherically-symmetric stationary states}

For a spherically-symmetric stationary state, there is no loss of
generality in assuming that the wave-function is real. The time-independent SN
equations can therefore be written in the form
\begin{eqnarray}
(r\psi)_{rr} &=& -r\psi U,\label{SSNE} \\
(rU)_{rr} &=& -r\psi^2, \nonumber
\end{eqnarray}
where we have introduced the variable $U=E-\phi$ and used a subscript
$r$ 
for $d/dr$.

The boundary conditions are that $\psi \rightarrow 0$ as 
$r \rightarrow
\infty$ and $\psi_r = 0 = \phi_r$ at $r=0$. If $(\psi,U,r)$ 
is a solution then so is
$(\lambda^2\psi,\lambda^2U,\lambda^{-1}r)$ for any nonzero $\lambda$. This
scaling can be used to impose the normalisation (\ref{I})
retrospectively. To solve (\ref{SSNE}), Moroz et al \cite{cq98} use a
shooting method. They set $U(0)=1$, which can be done with a suitable
$\lambda$ leaving the solution not correctly normalised,
then they choose values of $\psi(0)$ and integrate away from $r=0$. The
integration continues until the  
solution for $\psi$ diverges at some value of $r$. This value of $r$
can be pushed out to larger distances and the solution may
diverge to large positive or large negative values. By 
adjusting $\psi(0)$ it is possible to bracket a value at which
$\psi$ remains finite and then $E$ can be
identified from the limit of $U$ at large distances.

\begin{figure} [ht]
\includegraphics[height = 0.4\textheight, width=1\textwidth]{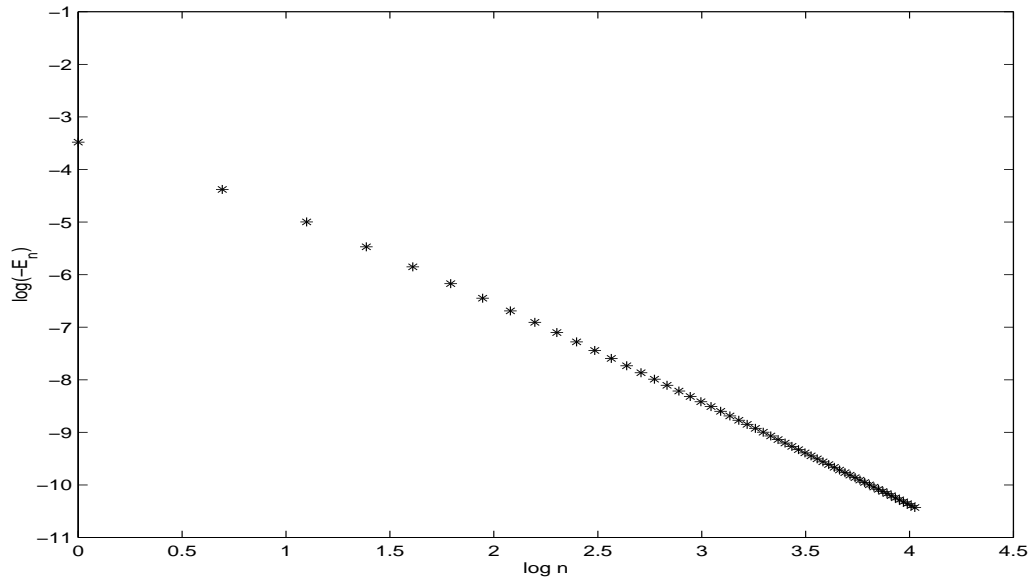}
\caption{Log-log plot of the energy values against $n$}
\label{log}
\end{figure}

With better triggering to recognise the diverging solutions, this 
technique can be refined and it becomes feasible to obtain the first 50 energy
levels in a reasonable time. The first 20 eigenvalues calculated 
with the above routine are given
in Table~\ref{e20}. The first 16 agree to this order with 
those given by Bernstein et al \cite{mp98}.
\begin{table} [ht]
\begin{center}
\begin{tabular}{|c|r|}
\hline
Number of zeros   & Energy Eigenvalue\\
\hline  
                  0  & -0.163\\
                  1  & -0.0308\\
                  2  & -0.0125\\
                  3  & -0.00675\\
                  4  & -0.00421\\
                  5  &-0.00287\\
                  6  & -0.00209\\
                  7  & -0.00158\\
                  8  & -0.00124\\
                  9  & -0.00100\\
                 10  & -0.000823\\
                 11  & -0.000689\\
                 12  & -0.000585\\
                 13  & -0.000503\\
                 14  & -0.000437\\
                 15  & -0.000384\\
                 16  &-0.000339 \\
                 17  &-0.000302 \\
                 18  &-0.000271 \\
                 19  &-0.000244 \\
                 20  &-0.000221 \\
\hline
\end{tabular}
\end{center}
\caption{ The first 20 eigenvalues}
\label{e20}
\end{table} 
A log-log plot of the $n$-th energy $E_n$ against $n$ is shown 
in figure~\ref{log}. The slope is asymptotically very close to -2,
which would be the exact value for the Hydrogen atom (as was
previously noted by Bernstein et al \cite{mp98}).
Once one has the eigenvalues, one can alternatively use a spectral
method to find the eigenfunctions and corresponding potentials at the
Chebyshev points. These will be needed in Sections 3 and 4 (where the
definition of Chebyshev points is also given).

\section{The perturbation equations}

In this section we shall set up the linear stability
problem, ready for numerical solution in Section 4. We look for a
solution to (\ref{ndimS}) which can be 
expanded in terms of a small
parameter $\epsilon$ as
\begin{eqnarray}
\psi({\bf{x}},t) &=& \psi_0 + \epsilon\psi_1 + \ldots, \label{pe}
\\
\phi({\bf{x}},t) &=& \phi_0 + \epsilon\phi_1 + \ldots, \nonumber
\end{eqnarray}
Substituting (\ref{pe}) into the SN equations (\ref{ndimS}) and 
equating powers of $\epsilon$ we obtain
\begin{eqnarray}
i{}\psi_{0t} + \nabla^2\psi_0 &=& \psi_0\phi_0 \\
\nabla^2\phi_0 &=& \vert\psi_0\vert^2 \nonumber
\label{oe0}
\end{eqnarray}
\begin{eqnarray}
i{}\psi_{1t} + \nabla^2\psi_1 &=& \psi_0\phi_1 + \psi_1\phi_0 \label{oe1}
\\
\nabla^2\phi_1 &=& \psi_0\bar{\psi}_1 + \bar{\psi}_0\psi_1. \nonumber
\end{eqnarray}
We restrict to spherical symmetry and take
\begin{eqnarray}
\psi_0 &=& R_0(r)e^{-i{}Et}, \\
\phi_0 &=& E - U_0(r),\nonumber
\end{eqnarray}
where $R_0$ and $U_0$, both real, are known from the work described in
Section 2. We seek
solutions of the form
\begin{eqnarray}
\psi_1 &=&\frac{1}{r} R(r,t)e^{-i{}Et},\label{R1} \\
\phi_1 &=& \frac{1}{r}\phi(r,t), \nonumber
\end{eqnarray}
where $R$ is complex and $\phi$ is real. With this choice of
variables, (\ref{oe1}) becomes
\begin{eqnarray}
i{}R_t + R_{rr} &=& R_0\phi - U_0R, \label{fop1}\\
\phi_{rr} &=& R_0\bar{R} + \bar{R}_0R.\nonumber
\end{eqnarray}
We look for a solution of the form
\begin{eqnarray}
R &=& (A+B)e^{\lambda{}t} + (\bar{A}-\bar{B})e^{\bar{\lambda}t},\label{eper} \\
\phi &=& We^{\lambda{}t} + \bar{W}e^{\bar{\lambda}t}, \nonumber
\end{eqnarray}
where we assume for now that $\lambda$ is not real and that $A, B$ and
$W$ are functions only of $r$. Now we substitute into 
(\ref{fop1}) and equate coefficients of
$\displaystyle{e^{\lambda{}t}}$ and 
$\displaystyle{e^{\bar{\lambda}t}}$ still assuming that $\lambda$ is not
real. This gives
\begin{eqnarray}
W_{rr} &=& 2R_0A,  \label{perta}\\
A_{rr} + U_0A - R_0W & = & -i\lambda{}B, \nonumber\\
B_{rr} + U_0B & = & -i\lambda{}A. \nonumber
\end{eqnarray}
If $(A,B,W)$ is a solution of 
(\ref{perta}) with eigenvalue $\lambda$ then
$(A,-B,W)$ is a solution with eigenvalue $-\lambda$. Thus eigenvalues
come in quadruples $(\lambda,-\lambda,\bar{\lambda}, -\bar{\lambda})$
for complex $\lambda$. If there exists a solution of 
(\ref{perta}) with
real $\lambda$ then the real and imaginary parts of the system
decouple. We may assume without loss of generality that $A$ and $W$
are real and $B$ is pure imaginary and then there is once again 
a solution of the
form of (\ref{eper}) to (\ref{fop1}). Thus
(\ref{perta}) gives the
perturbation equations with no restriction on $\lambda$. From the 
definitions (\ref{R1}) we need $A,B$ and $W$ to vanish at $r=0$
and to vanish asymptotically as $r$ tends to infinity (or at the outer
edge of the grid in a numerical calculation).\\

We note that $\lambda = 0$ will be an eigenvalue of 
(\ref{perta}) with 
$A = W = 0$, $B = R_0$. This is a trivial `zero-mode' corresponding to 
an infinitesimal constant rotation in the phase of the unperturbed
state. It will be found by
the numerical calculation in Section 5 and we shall discard it by hand.\\

We can obtain a condition on $\lambda^2$ as follows: multiply 
the second of (\ref{perta}) by $\bar{A}$ and the third by 
$\bar{B}$ and integrate to
find with the aid of the first that
\begin{eqnarray}
-i\lambda\int_0^\infty{}\bar{A}Bdr &=& 
\int_0^\infty{}(U_0|A|^2 - |A_r|^2 - \frac{1}{2}|W_r|^2)dr, \label{AB}\\
-i\lambda\int_0^\infty{}A\bar{B}dr &=& \int_0^\infty{}(U_0|B|^2 -
|B_r|^2)dr. \nonumber
\end{eqnarray}
The right-hand-sides in (\ref{AB}) are both real, so that the quotient
\begin{equation}
\frac{-i\lambda\int_0^\infty{}\bar{A}Bdr}
{i\bar{\lambda}\int_0^\infty{}\bar{A}Bdr},
\end{equation}
is real provided it is well-defined. Hence we have a dichotomy:
\begin{equation}
 either\,\, \lambda^2\,\, is\,\, real\,\, or \int_0^\infty{}\bar{A}Bdr = 0. 
\label{dich}
\end{equation}
This will provide a
useful check on the numerics in Section 5.

By a more sophisticated argument, which we shall relegate to an
appendix, we may obtain an upper bound on the real
part of any eigenvalue in terms of the energy of the state being
perturbed. For perturbations of a state with energy-eigenvalue $E$ this is
\begin{equation}
|\Re(\lambda)| \leq \frac{4}{9\pi^2}\sqrt{-E}.
\label{Relambda} 
\end{equation}
This is an interesting analytic result which limits the rate at which
an unstable state can decay. It also will provide a useful check in
Section 5. We now turn to the problem of solving the perturbation equations.

\section{Solving the perturbation equations}

 We solve the perturbation equations by a 
spectral method (see e.g. \cite{nt99}) which reduces the calculation
to that of solving a matrix eigenvalue problem. 

We choose a range $(0,L)$ for $r$, regarding $r=L$
as being at infinity for the purpose of setting boundary conditions, and set
$x=2\frac{r}{L}-1$. Next we choose a number $N$ and approximate any 
function $f(x)$ by the unique 
polynomial $p(x)$ of degree $N$ or less such that 
$p(x_k) = v_k$ for $k = 0,1 \ldots N$ , where $v_k = f(x_k)$ 
and $x_k = \cos(k\pi/N)$ are the Chebyshev points.

To approximate the derivative we introduce $w_k$ by 
$w_k = p'(x_k)$ for $k = 0,1 \ldots N$. Then
the differentiation matrix for polynomials of degree $N$, denoted $D_N$,
is defined by the equation
\begin{equation}
\left( \begin{array}{c} w_0 \\ w_1 \\ \vdots \\ w_N
\end{array} \right) = D_N \left( \begin{array}{c} v_0 \\ v_1 \\ \vdots
\\ v_N \end{array} \right).
\end{equation}
For an explicit formula for $D_N$, see e.g. \cite{nt99}. 

The second-derivative matrix is just $D_N^2$. The requirement that
$A$, $B$ and 
$W$ be zero at the boundary
is imposed by deleting the first and last rows and columns of the
relevant differentiation matrix, in this case $D_N^2$. This yields 
an $(N-1)\times(N-1)$ matrix conveniently denoted by $\tilde{D}_N^2$

For the perturbation equations (\ref{perta}) we therefore obtain the matrix 
eigenvalue equation 
\begin{equation}
\label{three}
 \left( \begin{array}{ccc} -2R_0 & 0 & \tilde{D}^2_N \\
0 & \tilde{D}^2_N+U_0 & 0 \\
-\tilde{D}^2_N-U_0 & 0 & R_0 \end{array} \right) \left(
\begin{array}{c}A\\ B \\ W \end{array} \right) = 
  i\lambda\left( \begin{array}{ccc} 0 & 0 & 0 \\
-I & 0 & 0\\
0 & I & 0 
\end{array} \right) \left( \begin{array}{c}A \\ B \\ W \end{array} \right),
\end{equation} 
where \\
\begin{equation}
\left( \begin{array}{c}A \\B\\ W \end{array} \right) 
= \left( \begin{array}{c}A(x_1) \\ \vdots \\ A(x_{N-1}) \\ B(x_1) \\
\vdots \\ B(x_{N-1}) \\ W(x_1) \\ \vdots \\ W(x_{N-1}) \end{array}
\right) ,
\end{equation}
and
\begin{eqnarray*}
R_0 & = & diag(R_0(x_1), R_0(x_2), \ldots, R_0(x_{N-1}))\\
U_0 & = & diag(U_0(x_1), U_0(x_2), \ldots, U_0(x_{N-1})). 
\end{eqnarray*}
where $diag$ indicates a diagonal matrix. 

One might worry that since the matrix on the right in (\ref{three}) 
is singular, accuracy in the eigenvalues was lost. 
By solving the equation for $W$ in terms of $A$ we can rewrite (\ref{three}) as
\begin{eqnarray}
     \left( \begin{array}{cc} 0 & \tilde{D}^2_N+U_0 \\
 \tilde{D}^2_N+U_0 -2R_0(\tilde{D}^{2}_N)^{-1}R_0 & 0 \end{array} \right) 
\left(
\begin{array}{c}A\\ B  \end{array} \right) \nonumber \\ 
 = -i\lambda\left( \begin{array}{cc} I & 0 \\
0& I  \end{array} \right) \left( \begin{array}{c}A \\ B  \end{array}
\right)
\label{altW}
\end{eqnarray}
However the difference between the
eigenvalues computed by using (\ref{altW}) and by using (\ref{three}) 
turns out to be small and we can just use
(\ref{three}).

\section{Numerical results}
We consider first the results obtained by solving (\ref{three})
about the 
ground state of the system (\ref{ntdSNa}),(\ref{ntdSNb}). Recall for comparison
that the 
eigenvalue for the ground state in
non-dimensionalized units is $-0.163$. The conversion factor to
dimensional units is given by (\ref{gamma}). 
\begin{figure} 
\includegraphics[height= 0.4\textheight, width=1\textwidth]{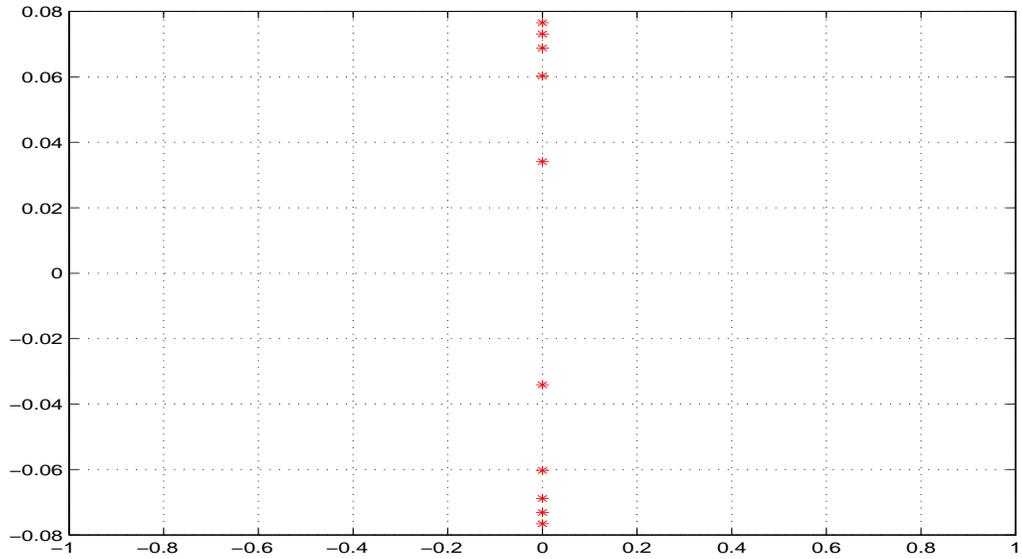}
\caption{The smallest eigenvalues of the perturbation about the ground
state}
\label{ss4p1}
\end{figure}
\begin{figure}
\includegraphics[height=0.4\textheight, width = 1\textwidth]{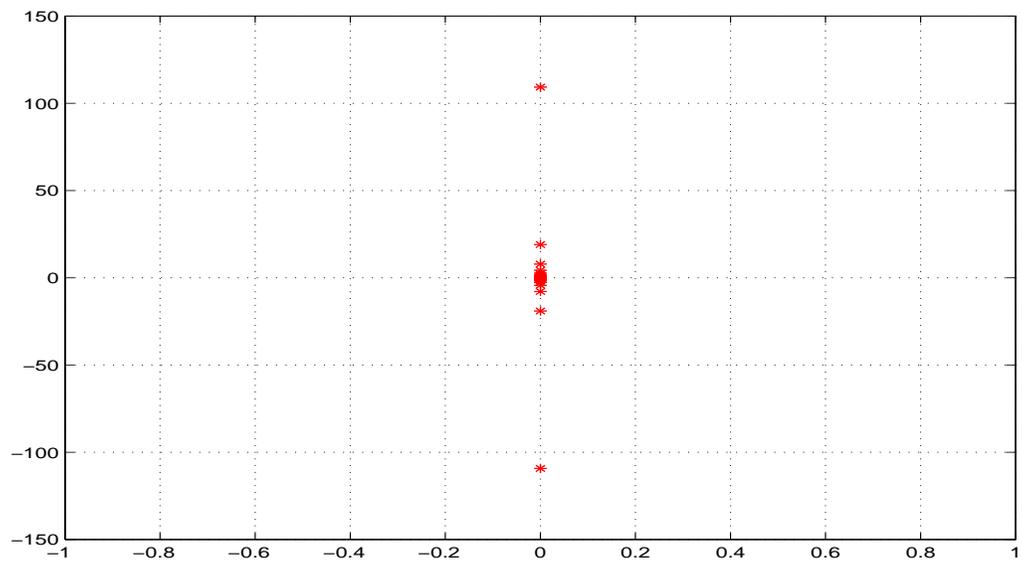}
\caption{All the computed eigenvalues of the perturbation about
the ground state}
\label{large}
\end{figure}
In figure~\ref{ss4p1} we plot those eigenvalues obtained by solving
(\ref{three}) which are closest to the origin, excluding 
the near-zero eigenvalue which does not
correspond to a non-trivial perturbation. To compute these results, 
we used $N = 60$ and $L = 150$. (That the near-zero eigenvalue 
found does indeed correspond to the
trivial zero-mode can be seen by plotting the corresponding 
eigenfunction. One finds  that
$A$, $W$ are very small $(O(10^{-7}))$ as compared to $B$, and that $B$ is
close to $R_0$, so that this solution indeed 
corresponds to the zero-mode foreseen in Section 3.)\\

In figure~\ref{large} we plot all the
eigenvalues obtained by solving (\ref{three}) (note the difference
in vertical scale from figure~\ref{ss4p1}) so that the quadruple of complex
eigenvalues appears as a pair) to show that up to 
these limits there are no eigenvalues other than imaginary ones. 
The first few eigenvalues obtained, now including the near-zero one, 
are presented in table~\ref{te1}. 
\begin{table}[htb]
\begin{center}
\begin{tabular}{|c|}
\hline
 $\pm{}0.000000116$ \\                    
 $\pm{}0.0341i$\\
 $\pm{}0.0603i$\\
 $\pm{}0.0688i$\\
 $\pm{}0.0731i$\\
 $\pm{}0.0765i$\\
 $\pm{}0.0810i$\\
 $\pm{}0.0867i$\\
\hline
\end{tabular}
\end{center}
\caption{Eigenvalues of the perturbation about the ground state}
\label{te1}
\end{table}
The non-trivial eigenvalues are all imaginary and we may conclude that the 
ground state is linearly stable.

\begin{figure}
\includegraphics[height = 0.4\textheight, width =
1\textwidth]{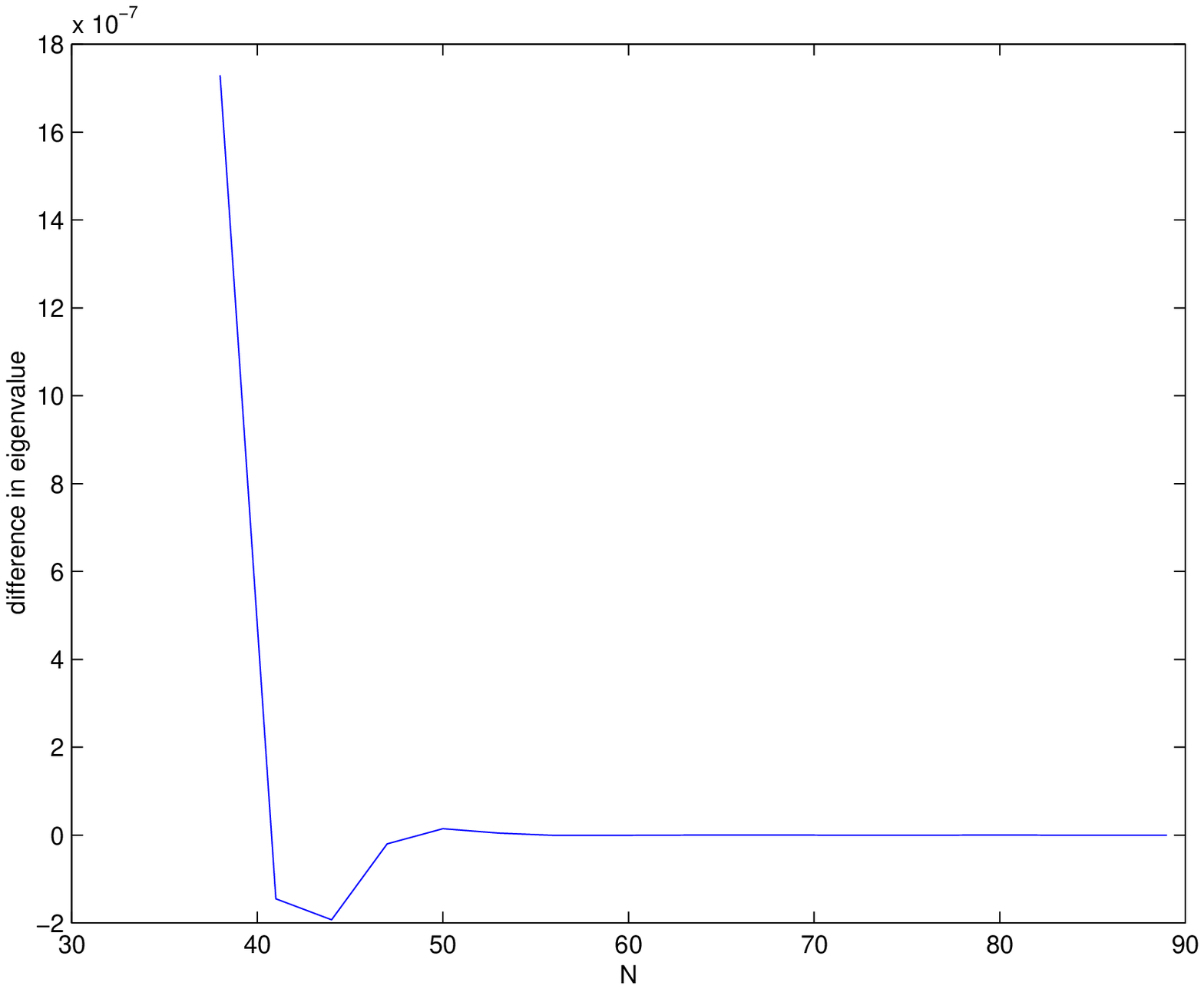}
\caption{The change in the sample eigenvalue with increasing values of $N$
($L = 150$)}
\label{ss4n1}
\end{figure}

\begin{figure}
\includegraphics[height = 0.4\textheight, width = 1\textwidth]{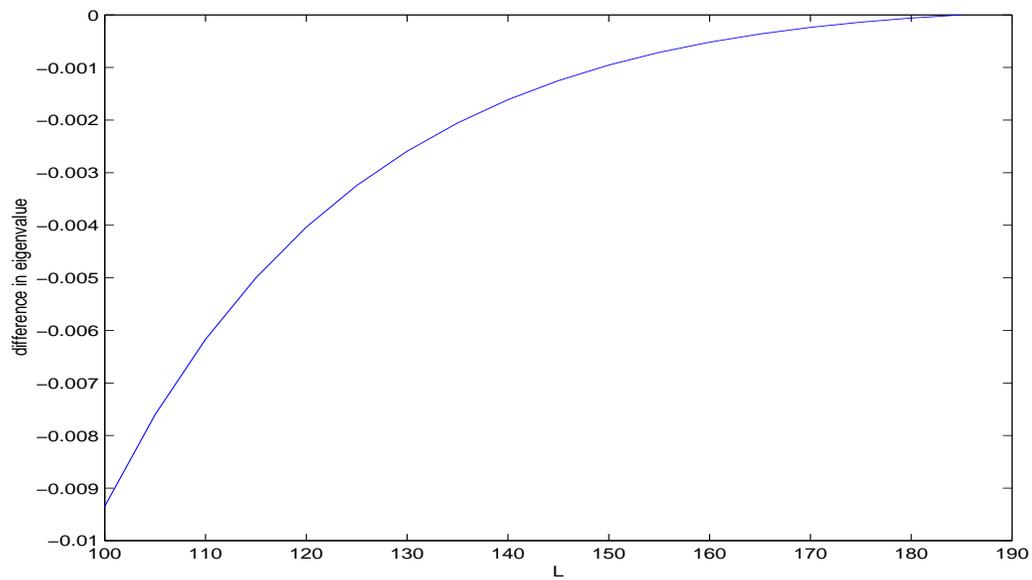}
\caption{The change in the sample eigenvalue with increasing values of $L$
($N = 60$)}
\label{ss4n2}
\end{figure}

\begin{figure}
\includegraphics[height= 0.4\textheight, width = 1\textwidth]{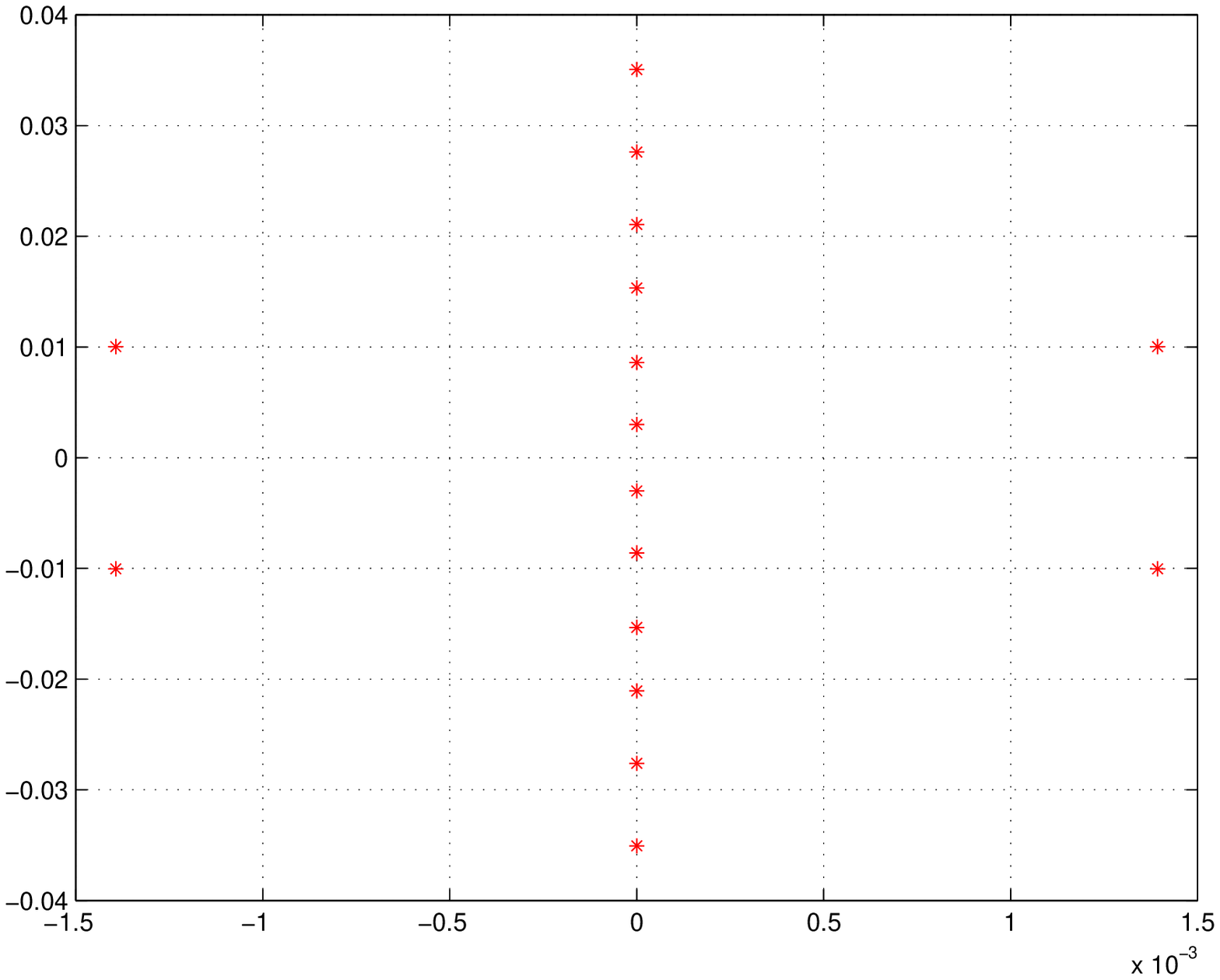}
\caption{The lowest eigenvalues of the perturbation about the second bound
state}
\label{SBS}
\end{figure}

\begin{figure}
\includegraphics[height= 0.4\textheight, width =
1\textwidth]{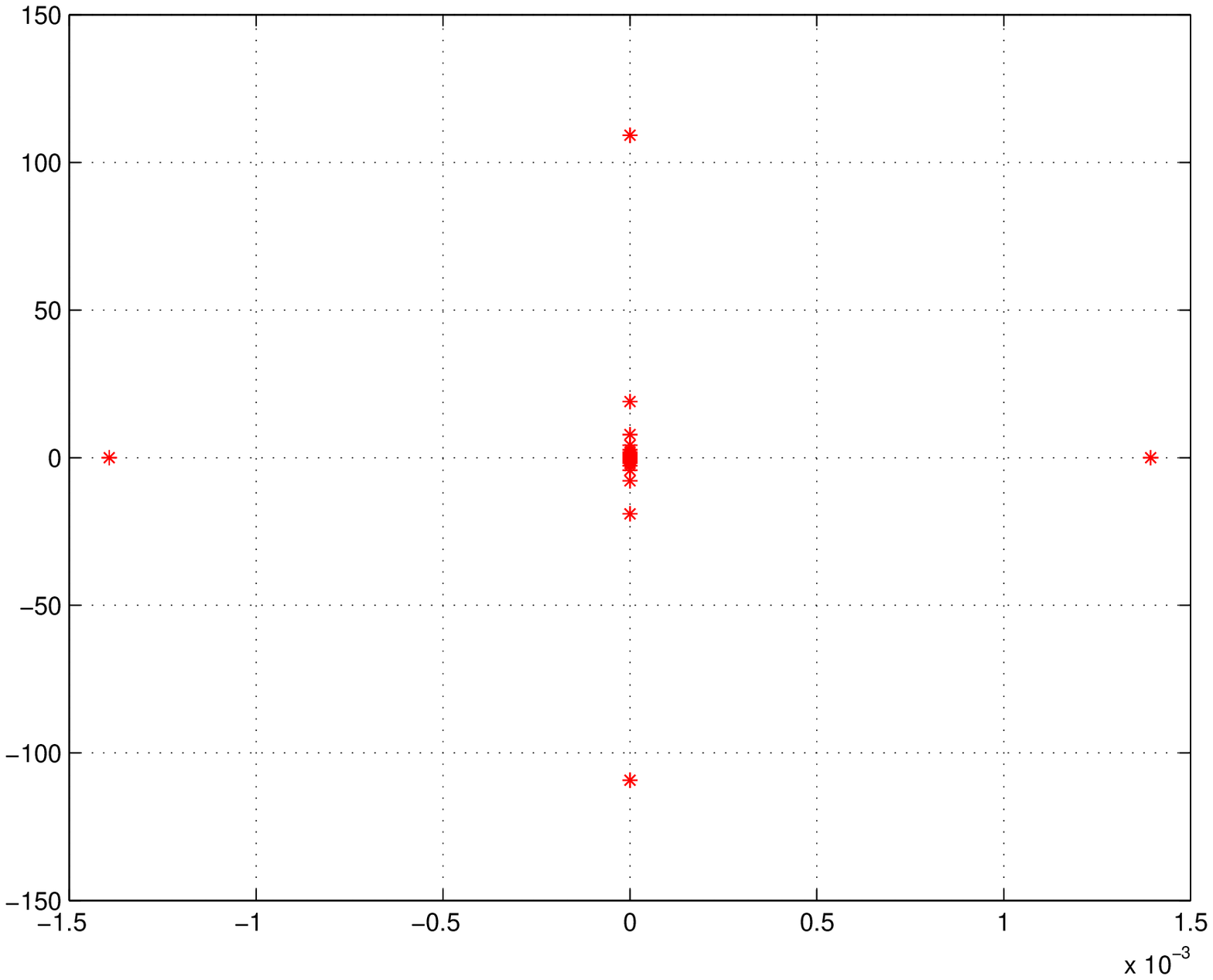}
\caption{All the computed eigenvalues of the perturbation about the
second bound state} 
\label{large2}
\end{figure}

To test convergence of the method  
we plot graphs of a calculated eigenvalue against increasing $N$ and $L$.
As an example we
show the graph of the fifth eigenvalue 
$0.0765i$
as the value of $N$ increases in figure~\ref{ss4n1} and as the value 
of $L$ increases in figure~\ref{ss4n2}. There is satisfactory convergence.\\
Using the same method we compute the numerical solutions of the
perturbation about the second state. This time we obtain some eigenvalues
with nonzero real parts. In figure~\ref{SBS} we plot the eigenvalues
nearest to the origin for 
the case where $N=60$ and $L =150$. 
In figure~\ref{large2} we plot all the eigenvalues obtained with a
different vertical scale (so that the quadruple of complex
eigenvalues appears as a pair) to see that up to these 
limits there are no other
complex ones.

From Section 3 we have two analytic conditions on the eigenvalues. The
first (\ref{dich}) relates to complex eigenvalues and we do now have
some. By the result of (\ref{dich})
we expect that for these $\int_0^\infty{}\bar{A}Bdr = 0$. To see
whether this is the case (up to numerical error) we compute
\begin{equation}
Q =
\frac{|\int_0^L{}\bar{A}Bdr|}
{(\int_0^L{}|A|^2dr)^\frac{1}{2}(\int_0^L{}|B|^2dr)^\frac{1}{2}}
\label{A*B}
\end{equation}
which we want to be much less than one. For $L =145$ and $N=60$ we
display in table~\ref{te2ab}
the calculated values of $Q$ with the eigenvalues nearest the origin. 
As expected, for the eigenvalues with nonzero real part, $Q$ is 
zero within numerical error and the result of (\ref{dich}) is confirmed.
\begin{table}[htb]
\begin{center}
\begin{tabular}{|c|c|}
\hline 
$\lambda $& $ Q$ \\
\hline
$                   0.003i$ & $0.235$\\
$                   0.00860i$ & $0.368$\\
$ -0.00139 - 0.010i$ & $3.53exp(-14)$\\
$  0.00139 + 0.010i$ & $3.92exp(-14)$\\
$                  0.0153i$ & $0.894$\\
\hline
\end{tabular}
\end{center}
\caption{$Q$ for different eigenvalues of the perturbation about the
second state}
\label{te2ab}
\end{table}

\begin{figure}
\includegraphics[height= 0.4\textheight, width= 1\textwidth]{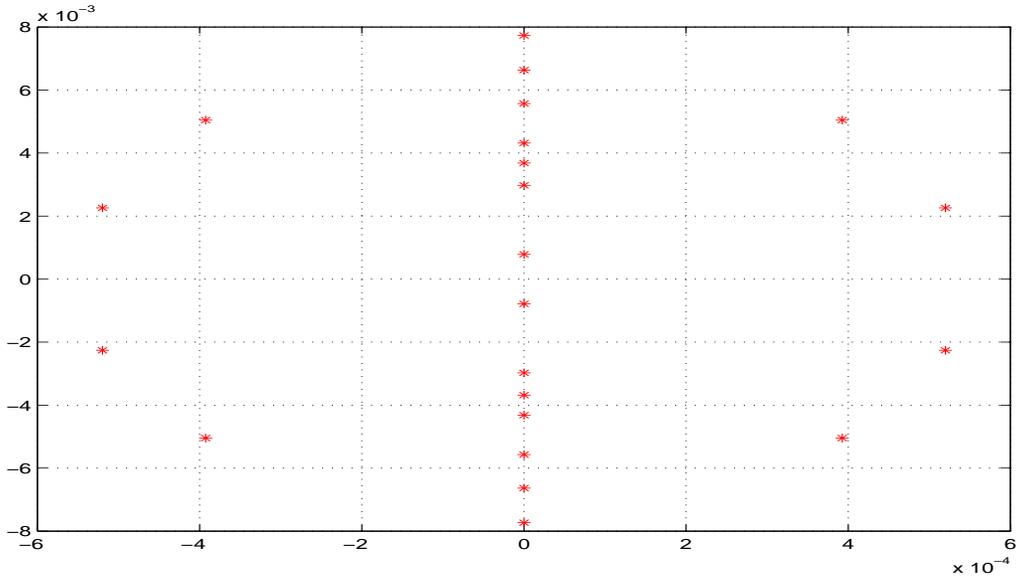}
\caption{The eigenvalues of the perturbation about the third bound
	state}
\label{T3B}
\end{figure}

For the third state, with $N = 60$ and $L = 450$, we present 
the eigenvalues in figure~\ref{T3B}. We have also calculated the
eigenvalues for the fourth state and a consistent pattern emerges:
there are $n$-quadruples of complex eigenvalues for perturbations
about the $(n+1)$-th bound state (equivalently for the $n$-th excited
state); the rest of the eigenvalues come in purely imaginary pairs.

With the computed eigenvalues for the higher states available we can
check their consistency with the
other result, (\ref{Relambda}), from Section 3. In table~\ref{tbre} we
compare the maximum of $\Re(\lambda)$ for the $(n+1)$-th state against
the bound (\ref{Relambda}). The computed
eigenvalues comply comfortably with this analytic bound.
\begin{table}[htb]
\begin{center}
\begin{tabular}{|c|c|c|c|}
\hline
State & Computed maximum $\Re(\lambda)$ & Bound from (\ref{Relambda})\\
\hline
 1 & 0               & 0.00362\\ 
 2 & 0.00139 & 0.00158\\ 
 3 & 0.000520 & 0.00101\\ 
 4 & 0.000225 & 0.000738\\ 
 5 & 0.000114 & 0.000583\\ 
 6 & 0.0000653 & 0.000482\\ 
\hline
\end{tabular}
\end{center}
\caption{Bound on the real part of the eigenvalues} 
\label{tbre}
\end{table}
When we have the eigenvalues for a particular state, we can check the
accuracy of the spectral method by solving (\ref{perta}) with each
eigenvalue in turn. Using a Runge-Kutta method, we arrive at the 
same eigenfunctions as by the spectral method.\\

To summarise, in this paper we have used a spectral method to analyse
 the linear stability of
the stationary solutions of the SN equations in the context of the
time-dependent SN equations. We have found that the ground-state 
solution is linearly stable
having purely imaginary eigenvalues and the $n$-th excited state has
$n$ unstable modes corresponding to $n$ quadruples of complex
eigenvalues, together with purely imaginary eigenvalues. These results
are tested for convergence and compliance with two analytical
results. They are also in line with the expectation one might have
from similar nonlinear problems like those described in \cite{volk}
and \cite{biz}. We shall consider the nonlinear stability in a second
paper where this picture is confirmed and one can see the excited
states decaying to the ground state.

\section*{Acknowledgement}
The work described in this paper formed part of the D.Phil. thesis of
the first author and he gratefully acknowledges the receipt of a grant
from EPSRC.

\section*{Appendix}
In this appendix, we shall prove (\ref{Relambda}) starting from
(\ref{perta}). The calculation uses the best constant for the Sobolev
inequality in three-dimensions given by Aubin (\cite{aub}). To work in
three-dimensions, we first define $a=A/r, b=B/r$ and $w=W/r$. In these
variables we can write (\ref{perta}) as
\begin{eqnarray}
\nabla^2w & = & 2R_0a, \label{app1} \\
\nabla^2a+U_0a-R_0w & = & -i\lambda b, \nonumber\\ 
\nabla^2b+U_0b & = & i\lambda a. \nonumber
\end{eqnarray}
We multiply the second of (\ref{app1}) by $\bar{a}$, the complex
conjugate of the third by $b$, add and integrate by parts to find
\begin{eqnarray}
i\int[ \lambda{}a\bar{a} + \bar{\lambda}b\bar{b}] =
-\int{}R_0b\bar{w}, 
\label{relam}
\end{eqnarray} 
where, as usual, the integrals are over $\bf{R}^3$.
Now we make repeated use of 
the Holder inequality and the Sobolev inequality in 3-dimensions:
\[
(\int F^6)^{\frac{1}{6}} \leq K(\int|\nabla F|^2)^{\frac{1}{2}}
\]
where $K$ is the relevant Sobolev constant, namely
\begin{equation}
K=\frac{2^{\frac{2}{3}}}{3^{\frac{1}{2}}\pi^{\frac{2}{3}}}.
\label{K}
\end{equation}
From the first of (\ref{app1}) and the Holder
inequality we find
\[
\int |\nabla w|^2 \leq 2\int|R_0aw| \leq 
2(\int|R_0|^3)^{\frac{1}{3}}(\int |a|^2)^{\frac{1}{2}}(\int
|w|^6)^{\frac{1}{6}}
\]
%
and using the Sobolev inequality in this gives
\begin{equation}
(\int|\nabla w|^2)^{\frac{1}{2}} 
\leq 2K(\int|R_0|^3)^{\frac{1}{3}}(\int |a|^2)^{\frac{1}{2}}.
\label{AW2}
\end{equation}
Next we calculate
\begin{eqnarray*}
\int |R_0bw| & \leq & (\int|R_0|^3)^{\frac{1}{3}}(\int |b|^2)^{\frac{1}{2}}
(\int |w|^6)^{\frac{1}{6}}\\ 
& \leq & K(\int|R_0|^3)^{\frac{1}{3}}(\int |b|^2)^{\frac{1}{2}}
(\int|\nabla w|^2)^{\frac{1}{2}}\\ 
& \leq & 2K^2(\int|R_0|^3)^{\frac{2}{3}}(\int |a|^2)^{\frac{1}{2}}
(\int |b|^2)^{\frac{1}{2}} 
\end{eqnarray*}
where the first step uses the Holder inequality, the second uses the Sobolev
inequality and the third uses equation (\ref{AW2}). We shall 
put this together with (\ref{relam}). 
First we normalise
 the perturbation so that 
\begin{equation}
\int|a|^2 = \cos^2\theta, \int|b|^2 = \sin^2\theta, 
\label{thet}
\end{equation}
for some $\theta$ and set $\lambda = \mu +i{}\nu$. Now from
(\ref{relam}) and (\ref{thet}) we have 
\[
|\mu + i{}\nu\cos{}2\theta| \leq
 K^2\sin{}2\theta(\int|R_0|^3)^{\frac{2}{3}}.
\]
%
from which it follows that
\[
|\mu| \leq  K^2(\int|R_0|^3)^{\frac{2}{3}}.
\]
We need an upper limit for the right-hand-side in this, which we
obtain as follows:
\begin{eqnarray*}
\int|R_0|^3 & \leq &
(\int R_0^2)^{\frac{3}{4}}(\int R_0^6)^{\frac{1}{6}}\\ 
& \leq & K^{\frac{3}{2}}(\int|\nabla R_0|^2)^{\frac{1}{2}}\\ 
& = &  K^{\frac{3}{2}}T^{\frac{3}{4}}
\end{eqnarray*}
using the Holder and Sobolev inequalities 
again and the definition (\ref{TV}) of $T$. Using
(\ref{TVE}) for $T$ in a stationary state and the definition (\ref{K}) of 
$K$ we finally arrive at the desired limit
\begin{eqnarray}
|\mu| \leq \frac{4}{9\pi^2}\sqrt{-E}. 
\end{eqnarray}

\end{document}